\documentclass[showpacs,preprintnumbers,amsmath,amssymb]{revtex4}
\usepackage{graphicx,epsf}
\def\be{\begin{equation}}
\def\ee{\end{equation}}
\def\bea{\begin{eqnarray}}
\def\eea{\end{eqnarray}}

\def\S{{\cal{S}}}

\def\dqa{{\delta q_\alpha}}
\def\dq{{\delta q}}
\begin{document}
\preprint{}
\title{Large-scale curvature and entropy perturbations for multiple
  interacting fluids}
\author{Karim A.~Malik$^1$, David Wands$^2$, and Carlo Ungarelli$^{2,3}$}
%
\affiliation{ 
$^1$GRECO, Institut d'Astrophysique de Paris, C.N.R.S.,
98bis Boulevard Arago, \\
75014 Paris, France\\
$^2$Institute of Cosmology and Gravitation, University of Portsmouth,
Portsmouth~PO1~2EG, United Kingdom\\
$^3$School of Physics and Astronomy, University of Birmingham,
Edgbaston, Birmingham~B15~2TT, United Kingdom}
%
\begin{abstract}
We present a gauge-invariant formalism to study the evolution of
curvature perturbations in a Friedmann-Robertson-Walker universe
filled by multiple interacting fluids. We resolve arbitrary
perturbations into adiabatic and entropy components and derive
their coupled evolution equations. We demonstrate that
perturbations obeying a generalised adiabatic condition remain
adiabatic in the large-scale limit, even when one includes energy
transfer between fluids. As a specific application we study the
recently proposed curvaton model, in which the curvaton decays
into radiation. We use the coupled evolution equations to show how
an initial isocurvature perturbation in the curvaton gives rise to
an adiabatic curvature perturbation after the curvaton decays.
\end{abstract}

\pacs{98.80.Cq \hfill Phys.\ Rev.\ D {\bf 67} (2003) 063516, 
PU-ICG-02/31, astro-ph/0211602v3}
\maketitle

\section{Introduction}

The primordial curvature perturbation plays a central role in
modern cosmology. It characterises large-scale density
perturbations in our Universe from which smaller scale structures
form via gravitational instability. Therefore much effort has been
devoted to understanding the evolution of the curvature
perturbation on large-scales in a general cosmology.
A gauge-invariant formalism for cosmological metric perturbations was
developed by Bardeen \cite{Bardeen} and the curvature perturbation (on
uniform density hypersurfaces) $\zeta$ was introduced by Bardeen et
al.~\cite{BST,Bardeen88} shortly afterwards as a convenient
gauge-invariant variable which remains constant for purely adiabatic
perturbations on large scales.
On large scales in an expanding universe  
it is essentially
equivalent to the comoving density perturbation
\cite{Lukash,Lyth1985,MFB}.

The constancy of the curvature perturbation $\zeta$ in the case of a
single perfect fluid follows directly from the local conservation of
the energy-momentum tensor, in a suitably defined large-scale limit
\cite{WMLL}. But $\zeta$ can change on arbitrarily large scales due to
a non-adiabatic pressure perturbation \cite{Mollerach,MFB,GBW,WMLL}.
Thus in a multi-fluid system it is in general necessary to follow the
coupled evolution of curvature and entropy (or isocurvature)
perturbations in order to determine the late-time curvature
perturbation.

There has been increasing interest in multi-field inflationary models
and the spectrum of curvature~\cite{Polarski1,Starobinsky,SS,GBW,MS}
and isocurvature~\cite{Polarski2,LM} perturbations that may be
produced and their correlations \cite{Langlois,chris}. In particular
it has recently been suggested that the large-scale curvature
perturbation $\zeta$ may be generated by initial isocurvature
perturbations in a ``curvaton'' field which subsequently decays into
radiation \cite{Enqvist,curvaton,MT,LUW}.
%


Kodama and Sasaki \cite{KS} developed a general formalism to
describe the evolution of cosmological perturbations with multiple
fluids (with corrections given in \cite{hama}). This formalism has
subsequently been used by a number of authors
\cite{KSrad1,KSrad2,hama,Hwang:fluids,Mar} (see also
\cite{bruni,nambu,dehnen}).
In particular Kodama and Sasaki applied their formalism to a
matter-radiation fluid in \cite{KSrad1,KSrad2}, where energy transfer
can be neglected.
One can argue on general grounds \cite{WMLL} that the entropy
perturbations evolve independently of the curvature perturbation on
large scales, but that the evolution of the large-scale curvature is
sourced by entropy perturbations.
Nonetheless there has been no detailed study of the evolution in
general of curvature and entropy perturbations including energy
transfer.
By contrast a formalism to study the coupled evolution equations for
curvature and entropy perturbations in models with multiple
interacting scalar fields has recently been developed by 
Gordon et al.~\cite{chris} and applied in a variety of scenarios
\cite{BMR,NR,Finelli,Ashcroft} (see also~\cite{Hwang:fields,nibbelink}).

In this paper we introduce a gauge-invariant formalism to follow the
coupled evolution of curvature and entropy perturbations in
multi-fluid cosmologies when energy transfer between fluids is
included. As an example we study the evolution of curvature and
entropy perturbations in a curvaton scenario where the decay of the
curvaton field represents the transfer of energy from the curvaton to
radiation. We compare the results of numerical solutions of the
coupled equations with analytic estimates based on the sudden decay
approximation, where the curvaton and radiation are assumed to be
non-interacting up until a given decay time.

\section{Governing equations}
\label{govsec}

In this section we give the governing equations for the general
case of an arbitrary number of interacting fluids in general
relativity. We will consider linear perturbations about a
spatially-flat FRW background model, as defined by the line
element
\be ds^2=-(1+2\phi)dt^2+2aB_{,i}dt dx^i
+a^2\left[(1-2\psi)\delta_{ij}+2E_{,ij}\right]dx^idx^j \,, \ee
where we use the notation of Ref.\cite{MFB} for the
gauge-dependent curvature perturbation, $\psi$, the lapse
function, $\phi$, and scalar shear, $\chi\equiv a^2\dot E - aB$.

Each fluid has an energy-momentum tensor $T^{\mu\nu}_{(\alpha)}$.
The total energy momentum tensor $T^{\mu\nu}=\sum_\alpha
T^{\mu\nu}_{(\alpha)}$, is covariantly conserved, but we allow for
energy transfer between the fluids,
\be
 \label{Qvector}
\nabla_\mu T^{\mu\nu}_{(\alpha)}=Q^\nu_{(\alpha)}\,,
\ee
where $Q^\nu_{(\alpha)}$ is the energy-momentum transfer to
the $\alpha$-fluid, which is subject to the constraint
\be
\label{Qconstraint}
\sum_\alpha Q^\nu_{(\alpha)}=0 \,.
\ee
The equations hold for any type of fluid, the only requirement
being the local conservation of the total energy-momentum tensor,
$\nabla_\mu T^{\mu\nu}=0$.

\subsection{Background equations}

The evolution of the background FRW universe is governed by the
Friedmann constraint
\bea
\label{Friedmann}
H^2 &=& \frac{8\pi G}{3}\rho \,,
\eea
and the continuity equation
\be
\label{continuity}
\dot\rho=-3H\left( \rho+P\right)\,,
\ee
where
the dot denotes a derivative with respect to coordinate time
$t$, $H\equiv \dot a/a$ is the Hubble parameter, and
$\rho$ and $P$ are the total energy density and the total
pressure
%
\be
\sum_\alpha \rho_{\alpha} =\rho \,, \qquad
\sum_\alpha P_{\alpha} =P \,.
\ee

The continuity equation for
each individual fluid in the background is thus \cite{KS}
\be
\dot\rho_{\alpha}
=-3H\left(\rho_{\alpha}+P_{\alpha}\right) +Q_{\alpha}\,,
\ee
where the energy transfer to the $\alpha$ fluid is given by
the time component of the energy-momentum transfer vector
$Q^0_{(\alpha)}=Q_{\alpha}$ in the background.
Equation~(\ref{Qconstraint}) implies that the background energy
transfer obeys the constraint
\be
\label{backconstraint}
\sum_\alpha Q_{\alpha}=0 \,.
\ee
%

\subsection{Perturbed equations}

Perturbing
the constraint equation (\ref{Friedmann}) yields the
first-order equation~\cite{KS,MFB,thesis}
\be \label{pertFriedmann}
3H\left(\dot\psi+H\phi\right)-\frac{\nabla^2}{a^2}
\left(\psi+H\chi\right) = -4 \pi G \delta\rho \,, \ee
where the comoving spatial Laplacian is denoted by $\nabla^2\equiv
\partial^2/\partial x^{i2}$, and the momentum constraint equation
(identically zero in the FRW background) is given by~\cite{KS,MFB,thesis}
\be
\dot \psi + H\phi =-4 \pi G \dq\,,
\ee
where $\delta\rho$ is the density perturbation and $\delta q$ the
scalar 3-momentum potential. 
%

Perturbing the continuity equation (\ref{continuity}) yields an evolution
equation for the total density perturbation
\cite{KS,thesis}
\be 
\label{pertcontinuity} \dot{\delta\rho} + 3H(\delta\rho +
\delta P) -\left(\rho+P\right) 3\dot\psi + \frac{\nabla^2}{a^2}
\left[ \dq+(\rho+P)\chi\right] = 0 \,,
\ee
while total momentum conservation is given by \cite{KS,thesis}
\be
\dot\dq+3H\dq+(\rho+P)\phi +\delta
P+\frac{2}{3}\frac{\nabla^2}{a^2}\Pi = 0\,,
\ee
where $\Pi$ is the total anisotropic stress.

The perturbed energy transfer vector, Eq.~(\ref{Qvector}),
including terms up to first order, is written as~\cite{KS}
\bea
 Q_{(\alpha)0} &=& -Q_{\alpha}(1+\phi) - \delta Q_\alpha
 \,,\\
 Q_{(\alpha)i}
  &=& \left( f_\alpha+\frac{Q_\alpha}{\rho+P}\dq\right)_{,i} \,,
\eea
and Eq.~(\ref{Qconstraint}) implies that the perturbed energy and
momentum transfer obey the constraints
\be
\label{pertconstraint}
\sum_\alpha \delta Q_{\alpha} = 0 \,,
\quad
\sum_\alpha f_\alpha =0\,.
\ee

The perturbed energy conservation equation for a particular fluid,
including energy transfer, is then given by
\be \label{pertenergyexact}
\dot{\delta\rho}_{\alpha}+3H(\delta\rho_{\alpha}+\delta P_{\alpha})
- \left(\rho_{\alpha}+P_{\alpha}\right)3\dot\psi
+ \frac{\nabla^2}{a^2}\left[ \dqa+(\rho_{\alpha}+P_{\alpha})\chi\right]
= Q_{\alpha}\phi+\delta Q_{\alpha}\,,
\ee
while the momentum conservation equation is
\be
\dot\dqa+3H\dqa+(\rho_\alpha+P_\alpha)\phi
+\delta P_\alpha+\frac{2}{3}\frac{\nabla^2}{a^2}\Pi_\alpha
= Q_\alpha \frac{\dq}{\rho+P} + f_\alpha\,,
\ee
where
the density, pressure, momentum and anisotropic stress
perturbations of the individual fluids are related to the total
density, pressure, momentum and anisotropic stress perturbations by
\be
 \sum_\alpha \delta\rho_{\alpha} =\delta\rho \,, \qquad
 \sum_\alpha \delta P_{\alpha} = \delta P \,, \qquad
 \sum_\alpha \dqa = \dq \,, \qquad
 \sum_\alpha \Pi_\alpha =\Pi\,.
\ee
%


\subsection{Gauge-invariant perturbations}

Both the density perturbation, $\delta\rho_\alpha$, and the
curvature perturbation, $\psi$, are in general gauge-dependent.
Specifically they depend upon the chosen time-slicing in an
inhomogeneous universe. However a gauge-invariant combination can
be constructed which describes the density perturbation on uniform
curvature slices or, equivalently the curvature of uniform density
slices.

The curvature perturbation on uniform total density hypersurfaces,
$\zeta$, is given by \cite{BST,WMLL}
\be
\label{zeta}
\zeta=-\psi-H\frac{\delta\rho}{\dot\rho} \,,
\ee
while the curvature perturbation on uniform $\alpha$-fluid density
hypersurfaces, $\zeta_{\alpha}$, is defined as\cite{WMLL}
\be
\label{zetaalpha}
\zeta_{\alpha}=-\psi-H\frac{\delta\rho_\alpha}{\dot\rho_\alpha} \,.
\ee

The total curvature perturbation (\ref{zeta}) is thus a weighted
sum of the individual perturbations
\be
\label{zetatot}
\zeta= \sum_\alpha
 \frac{\dot\rho_\alpha}{\dot\rho}  \zeta_\alpha \,,
\ee
while the difference between any two curvature perturbations
describes a relative entropy (or isocurvature) perturbation
 \be
 \label{defS}
  \S_{\alpha\beta}=3(\zeta_\alpha-\zeta_\beta)
   = -3H
\left(
  \frac{\delta\rho_\alpha}{\dot\rho_\alpha}
  - \frac{\delta\rho_\beta}{\dot\rho_\beta} \right) \, .
\ee
The classic example of just such a relative entropy perturbation
is a perturbation in the primordial baryon-photon ratio (with
negligible energy transfer between the two fluids)
\begin{equation}
 \S_{B\gamma} = 3(\zeta_B-\zeta_\gamma) =
 \frac{\delta\rho_B}{\rho_B} - \frac{3}{4}
 \frac{\delta\rho_\gamma}{\rho_\gamma} \,.
\end{equation}
This is also described as an initial isocurvature baryon density
perturbation as $\S_{B\gamma} \to\delta\rho_B/\rho_B$ in the limit
$\rho_B/\rho_\gamma\to0$.

{}From the definitions of the total curvature perturbation
(\ref{zetatot}) and the entropy perturbation (\ref{defS}), we get
\be
 \label{relation}
 \zeta_{\alpha}=\zeta+\frac{1}{3} \sum_\beta
  \frac{\dot\rho_\beta}{\dot\rho}\S_{\alpha\beta}
 \,.
\ee
%

\subsection{Long-wavelength limit}

To describe the evolution of long-wavelength perturbations we will
work in the `separate universes' picture~\cite{WMLL} where,
smoothing over sufficiently large scales, the universe looks
locally like an unperturbed (FRW) cosmology. Specifically we
assume that we can neglect the divergence of the momenta in the
zero-shear gauge, $\nabla^2(\delta
q_\alpha+(\rho_\alpha+P_\alpha)\chi)$, in
Eq.~(\ref{pertenergyexact}).

In this long-wavelength limit, the perturbed continuity
equation~(\ref{pertcontinuity}) becomes
\be
\dot{\delta\rho} + 3H(\delta\rho + \delta P)
= 3\left(\rho+P\right)\dot\psi \,.
\ee
Re-writing this equation in terms of the total curvature
perturbation, $\zeta$ in Eq.~(\ref{zeta}), gives~\cite{GBW,WMLL}
\be
\label{dotzetatot} 
\dot\zeta = -\frac{H}{\rho+P}\delta P_{\rm nad} \,,
\ee
where the non-adiabatic pressure perturbation is $\delta P_{\rm
nad}\equiv\delta P-c_{\rm{s}}^2\delta\rho$ and the adiabatic sound
speed is $c_{\rm{s}}^2=\dot{P}/\dot\rho$. Thus the total curvature
perturbation is constant on large scales for purely adiabatic
perturbations.

%
%
%
%

In the presence of more than one fluid,
the total non-adiabatic pressure perturbation, $\delta P_{\rm nad}$,
may be split into two parts,
\be
\label{deltaPnad}
\delta P_{\rm nad}\equiv \delta P_{\rm intr}+\delta P_{\rm rel}\,.
\ee
The first part is due to the intrinsic entropy perturbation of
each fluid
\be
\label{deltaPintr}
\delta P_{\rm intr}=\sum_\alpha \delta P_{\rm{intr},\alpha} \,,
\ee
where the intrinsic non-adiabatic pressure perturbation of each
fluid is given by
\be \label{deltaPintralpha} \delta P_{\rm{intr},\alpha}
\equiv \delta P_{\alpha} - c^2_{\alpha}\delta\rho_{\alpha} \,,
\ee
$c^2_{\alpha}\equiv \dot P_{\alpha}/ \dot\rho_{\alpha}$ is the
adiabatic sound speed of that fluid and the total adiabatic sound
speed
is the weighted sum of the adiabatic sound speeds of the individual
fluids,
%
\be
 c^2_{\rm{s}} =
  \sum_\alpha \frac{\dot\rho_\alpha}{\dot\rho} c^2_\alpha
  \,.
\ee
The second part of the non-adiabatic pressure perturbation
(\ref{deltaPnad}) is due to the {\em relative entropy
perturbation} between different fluids, denoted by
$\cal{S}_{\alpha\beta}$ in Eq.~(\ref{defS}),
\be
 \label{deltaPrel}
 \delta P_{\rm rel} \equiv
 \frac{1}{6H\dot\rho}
 \sum_{\alpha,\beta}\dot\rho_\alpha\dot\rho_\beta
\left(c^2_\alpha-c^2_\beta\right)\S_{\alpha\beta} \,.
 \ee

%

The time-dependence of the intrinsic entropy perturbation, $\delta
P_{{\rm intr},\alpha}$, of each fluid must be specified according to
the detailed modelling of that fluid.  For instance, if the fluid has
a definite equation of state $P_\alpha=P_\alpha(\rho_\alpha)$ then the
intrinsic non-adiabatic pressure perturbation vanishes.

The evolution of the relative entropy perturbation, $\cal{S}_{\alpha\beta}$,
follows from the time dependence of the individual curvature
perturbations $\zeta_\alpha$ and $\zeta_\beta$.
Equation~(\ref{pertenergyexact}) for the evolution of the
gauge-dependent density perturbations in the long-wavelength limit
reduces to
\be \label{pertenergy}
\dot{\delta\rho}_{\alpha}+3H(\delta\rho_{\alpha}+\delta
P_{\alpha}) = 3\left(\rho_{\alpha}+P_{\alpha}\right)\dot\psi
+Q_{\alpha}\phi+\delta Q_{\alpha}\,. \ee
Re-writing this in terms of the gauge-invariant curvature
perturbation $\zeta_\alpha$ defined in (\ref{zetaalpha}) gives
an evolution equation for the curvature perturbation on uniform
$\alpha$-fluid density hypersurfaces,
\begin{eqnarray}
 \label{dotzetaalpha}
\dot\zeta_\alpha &=&
 \frac{3H^2}{\dot\rho_\alpha}
 \left[ \delta P_{\alpha} - c^2_{\alpha}\delta\rho_{\alpha}
 \right]
 -{HQ_{\alpha}\over\dot\rho_\alpha} \left[ {\delta Q_\alpha\over
 Q_\alpha} + \left( {\dot\rho\over2\rho} -
   {\dot{Q}_\alpha\over Q_{\alpha}} \right)
   {\delta\rho_\alpha\over\dot\rho_\alpha}
  + H^{-1}\dot\psi + \phi \right] \nonumber\\
&=& \frac{3H^2\delta
  P_{\rm{intr},\alpha}}{\dot\rho_\alpha}
 -{H\delta Q_{\rm{nad},\alpha}\over\dot\rho_\alpha} \,.
\end{eqnarray}
For non-interacting, perfect fluids ($Q_\alpha=0$ and $\delta
P_{{\rm intr},\alpha}=0$) we have $\dot\zeta_\alpha=0$ and the
individual curvature perturbations for each fluid remain constant
in the long-wavelength limit~\cite{WMLL}.
But in general, the curvature perturbation $\zeta_\alpha$ may
change with time either due to the intrinsic non-adiabatic
pressure perturbation, $\delta P_{{\rm intr},\alpha}$ in
Eq.~(\ref{deltaPintralpha}) or due to what we will call the
`non-adiabatic' energy transfer, $\delta Q_{\rm{nad},\alpha}$.

Analogously to the total non-adiabatic pressure perturbation
(\ref{deltaPnad}), we will split the non-adiabatic energy transfer
into two parts,
\be
\delta Q_{{\rm nad},\alpha}\equiv
\delta Q_{{\rm intr},\alpha}+\delta Q_{{\rm rel},\alpha} \,.
\ee
The first part is the instrinsic non-adiabatic energy transfer
perturbations, defined as
\be
\label{deltaQintralpha}
\delta Q_{{\rm intr},\alpha} \equiv \delta Q_\alpha -
{\dot{Q}_\alpha\over\dot\rho_\alpha} \delta\rho_\alpha \,.
\ee
This is automatically zero if the local energy transfer $Q_\alpha$
is a function of the local density $\rho_\alpha$ so that $\delta
Q_\alpha = \dot{Q}_\alpha\delta\rho_\alpha/\dot\rho_\alpha$, just
as the intrinsic non-adiabatic pressure perturbation
(\ref{deltaPintralpha}) vanishes when $\delta P_\alpha =
\dot{P}_\alpha\delta\rho_\alpha/\dot\rho_\alpha$.
The second part is the relative non-adiabatic energy transfer
%
\be
\label{deltaQrelalpha}
\delta Q_{{\rm rel},\alpha} =
{Q_\alpha\dot\rho \over 2\rho} \left(
{\delta\rho_\alpha\over\dot\rho_\alpha} - {\delta\rho\over\dot\rho}
\right)
= - {Q_\alpha \over 6H\rho} \sum_\beta \dot\rho_\beta \S_{\alpha\beta}\,,
\ee
where we have used the background Einstein equations
(\ref{Friedmann}) and the perturbed Friedmann constraint equation
(\ref{pertFriedmann}) on large scales,
\be
 \label{einsteinconstraint} \dot\psi+H\phi =
  -\frac{H}{2}\frac{\delta\rho}{\rho} \,,
\ee
in order to write $\delta Q_{{\rm rel},\alpha}$ explicitly in terms of
the relative entropy perturbation, $\S_{\alpha\beta}$.

Note that the relative non-adiabatic pressure perturbation,
defined in Eq.~(\ref{deltaPrel}) is related to the relative
non-adiabatic energy transfer perturbation defined in
Eq.~(\ref{deltaQrelalpha}) as
\be
\delta P_{\rm{rel}}=-2\frac{\rho}{\dot\rho}\sum_\alpha
\frac{\dot\rho_{\alpha} c^2_{\alpha}}{Q_\alpha}
\delta Q_{{\rm rel},\alpha}\,.
\ee

%

The non-adiabatic pressure perturbations, $\delta
P_{\rm{intr},\alpha}$ and $\delta P_{\rm rel}$, and the
non-adiabatic energy transfers, $\delta Q_{\rm{intr},\alpha}$ and
$\delta Q_{\rm{rel},\alpha}$, are all automatically
gauge-invariant.
The intrinsic entropy perturbations, $\delta P_{\rm{intr},\alpha}$
and $\delta Q_{\rm{intr},\alpha}$, are both zero if the pressure
and local energy transfer are determined by the local energy
density. But even if the intrinsic entropy perturbations vanish,
there may be a non-adiabatic energy transfer due to the relative
entropy perturbation, $\zeta-\zeta_\alpha$. We interpret this as
due to a gravitational redshift (time-dilation) which perturbs the
rate of energy transfer with respect to coordinate time if the
uniform $\alpha$-density hypersurface does not coincide with the
uniform total density hypersurface, $\zeta_\alpha\neq\zeta$.

%


By taking the difference between the evolution equations
(\ref{dotzetaalpha}) for two fluids we obtain an evolution
equation for the relative entropy perturbation on large scales
\be
 \label{S_evol}
 \dot \S_{\alpha\beta} = 3H \left(
\frac{3H\delta P_{\rm{intr},\alpha} - \delta
  Q_{\rm{intr},\alpha}}{\dot\rho_\alpha}
-\frac{3H\delta P_{\rm{intr},\beta} - \delta
  Q_{\rm{intr},\beta}}{\dot\rho_\beta}\right)
+\sum_\gamma {\dot\rho_\gamma\over2\rho} \left(
  {Q_\alpha\over\dot\rho_\alpha} \S_{\alpha\gamma} -
  {Q_\beta\over\dot\rho_\beta} \S_{\beta\gamma} \right)
 \,.
\ee
%

Thus we see that any relative entropy perturbation
$\S_{\alpha\beta}$ is sourced on large scales only by intrinsic
entropy perturbations in the $\alpha$ and $\beta$ fluid, or by
other relative entropy perturbations. There is no source term
coming from the overall curvature perturbation and so adiabatic
perturbations (with no intrinsic or relative entropy perturbation)
remain adiabatic on large-scales even when one considers
interacting fluids.

\section{Curvaton decay}
\label{decaysec}

Having established a general formalism in which to study the evolution
of large-scale curvature and entropy perturbations including entropy
transfer between multiple fluids, we now study the specific case of a
non-relativistic matter decaying into radiation. In particular this
can be used to describe the decay of a massive curvaton field into
radiation~\cite{Enqvist,curvaton,MT,LUW}.

The curvaton scenario has recently been
proposed~~\cite{Enqvist,curvaton,MT,LUW} as a mechanism by which a
large-scale curvature perturbation can be produced from an initially
isocurvature perturbation.  If the curvaton is a light scalar field
(with mass less than the Hubble rate) the field may acquire an almost
scale-invariant spectrum of perturbations, $\zeta_\sigma$.
In the curvaton scenario radiation, $\rho_\gamma$, is supposed to
dominate the initial energy density after inflation and this is
assumed to be unperturbed, $\zeta_\gamma=0$.
Thus the curvaton perturbation is initially an isocurvature density
perturbation ($\zeta\simeq0$ and $\S_{\sigma\gamma}=3\zeta_\sigma$) and
remains an isocurvature perturbation while the relative density of the
curvaton remains negligible. However, once the Hubble rate drops below
the mass of the curvaton, the field begins to oscillate.  Averaged
over several oscillations the effective equation of state is $\langle
P_\sigma/\rho_\sigma \rangle=0$, i.e., the coherent oscillations of
the field are equivalent to a fluid of non-relativistic particles
\cite{Turner}.
As the energy density of non-relativistic particles grows
relative to the energy density of radiation, what was once an
isocurvature perturbation becomes a perturbation in the total
curvature, Eq.~(\ref{zetatot}).

Assuming the curvaton is unstable and decays into light particles
(``radiation'') with a decay rate $\Gamma$, this represents an energy
transfer from the pressureless curvaton fluid to the radiation fluid.
The precise amplitude of the resulting curvature perturbation,
relative to the initial curvaton perturbation, depends upon both the
initial density and the decay rate of the curvaton. We present the
equations for the evolution of the curvature and relative entropy
perturbations and solve them numerically, comparing with an analytic
approximation assuming an instantaneous decay.

\subsection{Background solution}

The energy transfer from the massive curvaton to light radiation
is described by
\bea
\label{defbackQa}
Q_{\sigma} &=& -\Gamma\rho_{\sigma} \,, \\
\label{defbackQb}
Q_{\gamma} &=& \Gamma\rho_{\sigma} \,,
\eea
where $\Gamma$ is the decay rate of the curvaton into
radiation, which we take to be a constant.
The energy conservation equations are therefore
\bea
\label{dotrhos}
\dot\rho_\sigma &=& -\rho_\sigma\left(3H+\Gamma\right)\,, \\
\label{dotrhog}
\dot\rho_\gamma &=& -4H\rho_\gamma+\Gamma\rho_\sigma\,, \\
\label{dotrhotot}
\dot\rho&=&-H\left(3\rho_\sigma+4\rho_\gamma\right) \,,
 \eea
where the Hubble expansion is given by
\be
\label{Friedmannmodel}
H^2 = {8\pi G\over3}\left( \rho_\sigma + \rho_\gamma \right) \,.
\ee

In order to solve the system of equations above numerically, it is
convenient to work in terms of the dimensionless density
parameters
\be
\Omega_\sigma \equiv \frac{\rho_\sigma}{\rho}\,,\quad
\Omega_\gamma \equiv \frac{\rho_{\gamma}}{\rho}\,,
\ee
and the dimensionless ``reduced'' decay rate
\be
\quad g\equiv\frac{\Gamma}{\Gamma+H} \,,
\ee
which varies monotonically from $0$ to $1$ in an expanding
universe.

\begin{figure}
\begin{center}
\includegraphics[angle=-90, width=70mm]{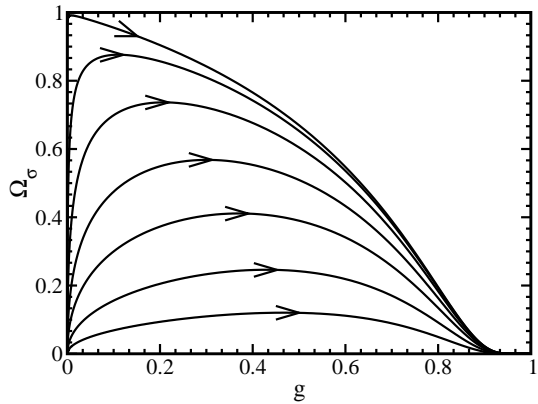} \\
\vspace{0.2cm}
\caption[phaseplot]{\label{phaseplot} Phase-plane showing
trajectories for the background solutions in the curvaton model in
Eqs.(\ref{din3}--\ref{con1}).}
\end{center}
\end{figure}

The background equations (\ref{dotrhos}-\ref{Friedmannmodel}) can
then be written as an autonomous system
\bea
\label{omegasprime}
 \Omega_\sigma' &=& \Omega_\sigma
 \left(\Omega_\gamma-{g\over1-g}\right) \label{din3} \,,\\
 \Omega_\gamma' &=& \Omega_\sigma
 \left({g\over1-g}-\Omega_\gamma\right) \label{din1}\,,\\
\label{gprime}
 g' &=& {1\over2}(4-\Omega_\sigma)(1-g)g \label{din4}\,,
\eea
%
%
where $^{\prime}$ denotes differentiation with respect to the
number of e-foldings $N\equiv \ln a$.
The density parameters are subject to the Friedmann constraint
(\ref{Friedmannmodel}) which requires
\be
 \Omega_\sigma + \Omega_\gamma = 1 \label{con1}\,.
\ee
There are only two independent dynamical equations and the generic
solutions follow trajectories in a compact two-dimensional phase-plane
($0\leq g\leq1$, $0\leq \Omega_\sigma\leq1$), illustrated in
Figure~\ref{phaseplot}.
%

The dynamical system (\ref{din3}--\ref{con1}) admits three fixed
points
\begin{description}
 \item[{\it (A)}] $\Omega_\gamma=1$, $\Omega_\sigma=0$, $g=0$,
 \item[{\it (B)}] $\Omega_\gamma=0$, $\Omega_\sigma=1$, $g=0$,
 \item[{\it (C)}] $\Omega_\gamma=1$, $\Omega_\sigma=0$, $g=1$.
\end{description}
Generic solutions start at the unstable repellor $(A)$ and approach the
stable attractor $(C)$ at late times.  At early times
($\Omega_\gamma\simeq1$, $g\ll1$) we find
$g\propto\Omega_\sigma^2\propto a^2$.  The standard radiation
dominated cosmology corresponds to evolution along the line
$\Omega_\sigma=0$. However solutions can approach arbitrarily close 
the curvaton-dominated saddle point $(B)$ before the curvaton decays
and $\Omega_\sigma\to0$ once again.

\subsection{Perturbations}

Both the curvaton and radiation fluids have fixed equations of
state ($\delta P_\sigma=0$ and $\delta
P_\gamma=\delta\rho_\gamma/3$) and hence there can be no intrinsic
non-adiabatic pressure perturbation ($\delta P_{{\rm
intr},\sigma}=0$ and $\delta P_{{\rm intr},\gamma}=0$). However
the total curvature perturbation, $\zeta$, does change on large
scales in the presence of a relative entropy
perturbation~(\ref{defS})
\be
 \label{defSsigmagamma}
 \S_{\sigma\gamma} \equiv 3 (\zeta_\sigma - \zeta_\gamma)
 \,,
\ee
which leads to a non-adiabatic pressure
perturbation~(\ref{deltaPrel}).
The evolution of the total curvature perturbation $\zeta$, using
Eqs.~(\ref{dotzetatot}), is
\be \label{dotzetatotmodel}
 \dot\zeta = {H\over3}
 \frac{\dot\rho_{\sigma}\dot\rho_{\gamma}}{\dot\rho^2}
  \S_{\sigma\gamma}
 \,.
\ee

We assume that the curvaton decay rate $\Gamma$ is fixed by
microphysics (i.e., $\delta\Gamma=0$) and hence the perturbed
energy transfer is simply given by
\bea
\delta Q_{\sigma}&=&-\Gamma\delta\rho_{\sigma} \,, \\
\delta Q_{\gamma}&=&\Gamma\delta\rho_{\sigma} \,.
\eea
This energy transfer is determined solely by the local density of
the curvaton and hence there is no intrinsic non-adiabatic energy
transfer from the curvaton, $\delta Q_{\rm{intr},\sigma}=0$.
However the radiation suffers an intrinsically non-adiabatic
energy transfer from the curvaton decay (\ref{deltaQintralpha})
\begin{equation}
\delta Q_{\rm{intr},\gamma} = \Gamma \left( \delta\rho_\sigma -
{\dot\rho_\sigma\over\dot\rho_\gamma} \delta\rho_\gamma \right)\,,
\end{equation}
which is proportional to the relative entropy perturbation
(\ref{defSsigmagamma}) between the radiation and curvaton
\begin{equation}
\delta Q_{\rm{intr},\gamma} = -{\Gamma\over 3H} \dot\rho_\sigma
\S_{\sigma\gamma} \,.
\end{equation}
The relative non-adiabatic energy transfers (\ref{deltaQrelalpha})
are also non-zero and given by,
\bea
\delta Q_{\rm{rel},\sigma} &=&
-\frac{\Gamma\rho_{\sigma}\dot\rho}{2\rho}\left(
\frac{\delta\rho_{\sigma}}{\dot\rho_{\sigma}}-\frac{\delta\rho}{\dot\rho}
\right) \,, \\
\delta Q_{\rm{rel},\gamma}&=&
\frac{\Gamma\rho_{\sigma}\dot\rho}{2\rho}\left(
\frac{\delta\rho_{\gamma}}{\dot\rho_{\gamma}}-\frac{\delta\rho}{\dot\rho}
\right) \,,
\eea
which can be rewritten in terms of the relative entropy
perturbation as
\bea
 \delta Q_{\rm{rel},\sigma} &=&
 \frac{\Gamma\rho_{\sigma}}{6H\rho}
 \dot\rho_\gamma \S_{\sigma\gamma}\,, \\
 \delta Q_{\rm{rel},\gamma}&=&
 \frac{\Gamma\rho_{\sigma}}{6H\rho}
 \dot\rho_\sigma \S_{\sigma\gamma}\,.
\eea
Thus the evolution equations (\ref{dotzetaalpha}) for the
curvature perturbation on uniform curvaton density hypersurfaces,
$\zeta_\sigma$, and uniform radiation density hypersurfaces,
$\zeta_\gamma$, are given by
\bea
\label{dotzetacurv}
\dot\zeta_\sigma
&=&
-{\Gamma\over6} {\rho_\sigma\over\rho}
{\dot\rho_\gamma\over\dot\rho_\sigma}\S_{\sigma\gamma} \,, \\
\label{dotzetarad}
\dot\zeta_\gamma
&=&
{\Gamma\over3}
{\dot\rho_\sigma\over\dot\rho_\gamma}
\left(1-{\rho_\sigma\over2\rho}\right)
\S_{\sigma\gamma} \,. \eea
The evolution equation for the relative entropy perturbation
$S_{\sigma\gamma}$ is, from Eq.~(\ref{S_evol}), 
\be
 \label{dotS}
 \dot \S_{\sigma\gamma}
=
\frac{\Gamma}{2}
\frac{\dot\rho_{\sigma}}{\dot\rho_{\gamma}}\frac{\rho_\sigma}{\rho}
\left(1-\frac{\dot\rho_{\gamma}^2}{\dot\rho_{\sigma}^2} \right)
\S_{\sigma\gamma}  \,. \ee

Equations~(\ref{dotzetatotmodel}) and (\ref{dotS}) form a closed
system of first-order equations for the evolution of the adiabatic and
entropy perturbations, $\zeta$ and $\S_{\alpha\beta}$, on large-scales
in the curvaton model.  They clearly demonstrate the general 
principle that the total curvature perturbation, $\zeta$, evolves on 
large scales in the presence of a relative entropy perturbation, 
$\S_{\alpha\beta}$, while the entropy perturbation obeys a 
homogeneous evolution equation,
unaffected by the large-scale curvature perturbation.
Alternatively we could use Eqs.~(\ref{dotzetacurv}) and
(\ref{dotzetarad}) as a closed system of first-order equations for
$\zeta_\sigma$ and $\zeta_\gamma$,
remembering that $\S_{\sigma\gamma}=3(\zeta_\sigma-\zeta_\gamma)$.

However the evolution equation (\ref{dotzetarad}) for $\zeta_\gamma$
and, hence, the evolution equation (\ref{dotS}) for
$\S_{\sigma\gamma}$ both become singular whenever
$\Gamma\rho_\sigma=4H\rho_\gamma$ and $\dot\rho_\gamma=0$.  This is
due to the uniform $\rho_\gamma$ hypersurface becoming ill-defined
rather than any breakdown of perturbation theory on generic
hypersurfaces. In particular the uniform-$\rho_\sigma$ and uniform
total energy density hypersurfaces remain well-behaved.

In practice we will use the two non-singular evolution
equations (\ref{dotzetatotmodel}) and (\ref{dotzetacurv}) for
$\zeta$ and $\zeta_\sigma$, respectively. In terms of the
dimensionless background variables we have two coupled evolution
equations
\bea
\label{zetaprime}
\zeta^{\prime}
&=&
\frac{\Omega_\sigma(2g-3)}{(1-g)(4-\Omega_\sigma)}
(\zeta-\zeta_{\sigma}) \,, \\
\label{zetasigmaprime}
\zeta_{\sigma}^{\prime}
&=&
\frac{g(4-\Omega_\sigma)}{2(3-2g)}(\zeta-\zeta_{\sigma})
 \,.
 \eea
%

\label{numsec}

To calculate the final curvature perturbation on large scales produced
in the curvaton scenario we start with initial conditions close to the
point $(A)$ for the background variables ($g\ll1$,
$\Omega_\sigma\ll1$) and unperturbed radiation, but perturbed curvaton
fluid:
\bea
 \zeta_\gamma &=& 0 \,,\\
 \zeta_\sigma &=& \zeta_{\sigma,{\rm in}} \,.
\eea
{}From the definitions of the total curvature perturbation and the
entropy perturbation Eqs.~(\ref{zetatot}) and (\ref{defS}), this
corresponds to initial values
\bea
\zeta &=& {3\Omega_{\sigma,{\rm in}}\over 4-\Omega_{\sigma,{\rm in}}}
 \, \zeta_{\sigma,{\rm in}} \,,\\
\S_{\sigma\gamma} &=& 3 \, \zeta_{\sigma,{\rm in}} \,.
\eea
This is an initial isocurvature perturbation in the sense that
$\zeta\to0$ in the early time limit, $(A)$ , where
$\Omega_{\sigma,{\rm in}}\to0$.

Starting from these initial conditions we use Eqs.~(\ref{zetaprime}) and
(\ref{zetasigmaprime}) to follow the evolution of $\zeta$ and
$\zeta_\sigma$ until we reach the late time attractor $(C)$ where
$g\to1$ and $\Omega_\sigma\to0$. At late times the perturbations too
approach a fixed point attractor where $\zeta_\gamma=\zeta_\sigma$ and 
\bea
\label{defr}
\zeta &=& r \zeta_{\sigma,{\rm in}} \,,\\
\S_{\sigma\gamma} &=& 0 \,.
\eea
This is an adiabatic primordial perturbation, where
the final value of the large-scale curvature perturbation,
$\zeta$, is related to the initital curvaton perturbation
$\zeta_{\sigma,{\rm in}}$ by a parameter $r$~\cite{curvaton,LUW} which
is determined by the numerical solution of Eqs.~(\ref{omegasprime}),
(\ref{gprime}), (\ref{zetaprime}) and
(\ref{zetasigmaprime}).

Thus, we can represent the integrated effect upon the large-scale curvature
and entropy perturbations of the curvaton growth and decay by the
transfer matrix:
\be
 \label{transfer}
\left(
\begin{array}{c}
\zeta \\ \S_{\sigma\gamma}
\end{array}
\right)_{\rm out}
 =
\left(
\begin{array}{cc}
1 & r/3 \\
0 & 0
\end{array}
\right)
\left(
\begin{array}{c}
\zeta \\ \S_{\sigma\gamma}
\end{array}
\right)_{\rm in} \,.
\ee

\begin{figure}
\begin{center}
\includegraphics[angle=-90, width=70mm]{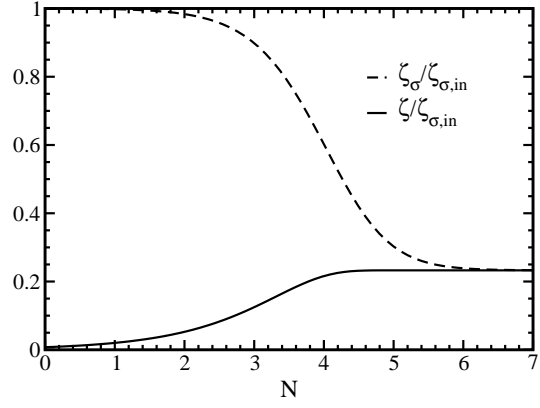} \\
\vspace{0.2cm}
\caption[pix]{\label{pix1} Evolution of the normalised
curvature perturbation on uniform curvaton density 
hypersurfaces, $\zeta_\sigma/\zeta_{\sigma,{\rm in}}$, and of
the normalised total curvature perturbation, 
$\zeta/\zeta_{\sigma,{\rm in}}$, as a function of the
number of e-foldingss, starting with 
$\zeta_\sigma/\zeta_{\sigma,{\rm in}}=1$ and initial
density and decay rate $\Omega_\sigma=10^{-2}$ and
$\Gamma/H=10^{-3}$. }
\end{center}
\end{figure}

\begin{figure}
\begin{center}
\includegraphics[angle=-90, width=70mm]{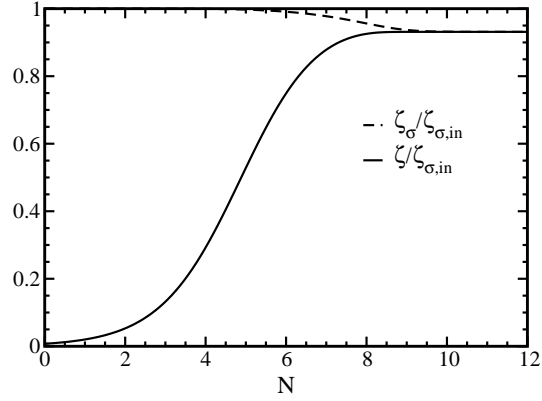} \\
\vspace{0.2cm}
\caption[pix]{\label{pix2} 
Evolution of the normalised curvature perturbation
on uniform curvaton density hypersurfaces, 
$\zeta_\sigma/\zeta_{\sigma,{\rm in}}$, and of
the normalised total curvature perturbation, 
$\zeta/\zeta_{\sigma,{\rm in}}$, as a function of the
number of e-foldingss, starting with 
$\zeta_\sigma/\zeta_{\sigma,{\rm in}}=1$ and initial
density and decay rate $\Omega_\sigma=10^{-2}$ and
$\Gamma/H=10^{-6}$. }
\end{center}
\end{figure}

\begin{figure}
\begin{center}
\includegraphics[width=80mm]{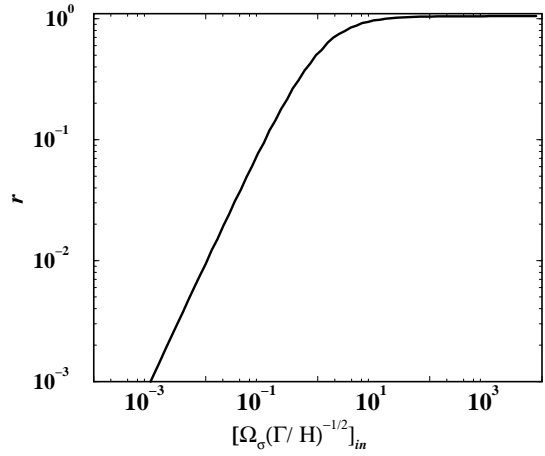} \\
\caption[pix]{\label{pix3} Transfer parameter $r$ defined in
  Eq.~(\ref{defr}) obtained from numerical solutions
  as a function of the initial value of
  $\Omega_\sigma/(\Gamma/H)^{1/2}$.} 
\end{center}
\end{figure}

Examples of the evolution of large-scale perturbations for two
different choices of initial conditions are shown in
Figures~\ref{pix1} and ~\ref{pix2}. The resulting value for the
transfer parameter $r$ defined in Eq.~(\ref{defr}) depends upon the
maximum value of $\Omega_{\sigma}$ before the curvaton decays. If the
curvaton dominates before it decays, i.e., $\Omega_{\sigma,{\rm
    dec}}\simeq1$, we have $r\simeq1$ as in the case shown in
Fig.\ref{pix2}. More generally, $r$ is a one-dimensional function of
the initial value of $\Omega_\sigma/(\Gamma/H)^{1/2}$ which determines
which trajectory is followed in the two-dimensional
$(g,\Omega_\sigma)$ phase-plane, Figure~\ref{phaseplot}. The precise
dependence of $r$ upon the initial value of
$\Omega_\sigma/(\Gamma/H)^{1/2}$ is shown in Figure~\ref{pix3}.

\subsection{Comparison with sudden decay approximation}
\label{suddsec}

Previous analyses \cite{curvaton,Bozza,LUW} have relied on the
assumpton of ``sudden decay'' to estimate the final curvature
perturbation produced after curvaton decay.
In this approximation the energy transfer
$Q_\sigma=-\Gamma\rho_\sigma$ is assumed to be negligible until the
decay time, defined by $\Gamma/H$ reaching some critical value
$\Gamma/H_{\rm dec}$ of order unity, at which time all the energy
density of the curvaton field is rapidly converted into radiation.

In the absence of energy transfer the individual curvature
perturbations $\zeta_\sigma$ and $\zeta_\gamma$, defined by
Eq.(\ref{zetaalpha}), remain constant on large scales [see
Eq.~(\ref{dotzetaalpha})]. Thus the total curvature perturbation,
Eq.~(\ref{zetatot}), is given by
\begin{equation}
 \zeta \approx f \zeta_\sigma + (1-f) \zeta_\gamma
\,,
 \end{equation}
where $\zeta_\sigma$ and $\zeta_\gamma$ are constant and the only
time dependence arises from the time dependence of the weight
given to the curvaton perturbation
\begin{equation}
\label{deff}
f \equiv {3\Omega_\sigma \over 3\Omega_\sigma+4\Omega_\gamma}\,.
\end{equation}

After the curvaton decays into radiation all the energy density in
the model has a unique equation of state, hence $\delta P_{\rm
nad}=0$ in Eq.~(\ref{dotzetatot}), and $\zeta$ becomes constant on
large scales. Hence in the curvaton scenario, where the initial
curvature perturbation $\zeta_\gamma$ is assumed to be negligible,
the resulting adiabatic curvature perturbation after curvaton
decay is given by
\begin{equation}
 \zeta_{\rm out} \approx f_{\rm dec} \zeta_{\sigma,{\rm in}} \,.
\end{equation}

In terms of an initial value for $\Omega_{\sigma,{\rm in}}$ and reduced
decay rate $\Gamma/H_{\rm in}$, we can write $f_{\rm dec}$ as
\begin{equation}
\label{fdec}
f_{\rm dec} \equiv {3\Omega_{\sigma,{\rm in}} \over
3\Omega_{\sigma,{\rm in}} + 4(1-\Omega_{\sigma,{\rm in}}) y_{\rm dec}}\,,
\end{equation}
where $y_{\rm dec}=a_{\rm in}/a_{\rm dec}$ is the ratio of initial scale factor to that
at decay. We can calculate this from the Friedmann equation for
non-interacting matter and radiation which can be written as
\begin{equation}
\left( {H\over H_{\rm in}} \right)^2 = (1-\Omega_{\sigma,{\rm in}}) \left( {a_{\rm in}\over
    a} \right)^4
+ \Omega_{\sigma,{\rm in}} \left( {a_{\rm in}\over a} \right)^3 \,.
\end{equation}
Thus the epoch of decay, $y_{\rm dec}$, is given by the one real root,
$0<y_{\rm dec}<1$ of
\begin{equation}
 \label{ystar}
(1-\Omega_{\sigma,{\rm in}}) y_{\rm dec}^4
 + \Omega_{\sigma,{\rm in}}\, y_{\rm dec}^3 -
\left( \frac{\Gamma/H_{\rm in}}{\Gamma/H_{\rm dec}} \right)^2 = 0 \,,
\end{equation}
and $f_{\rm dec}$ is then obtained from Eq.~(\ref{fdec}).

There are two limiting cases:
\begin{equation}
f_{\rm dec} \approx
\left\{
\begin{array}{ll}
{3\over4} \Omega_{\sigma,{\rm in}}
(1-\Omega_{\sigma,{\rm in}})^{-3/4}
\left( {\Gamma/H_{\rm dec}\over \Gamma/H_{\rm in}}
\right)^{1/2}
& \quad {\rm for} \ y_{\rm dec}\gg
\Omega_{\sigma,{\rm in}}/(1-\Omega_{\sigma,{\rm in}}) \\
1 - {4\over3} \Omega_{\sigma,{\rm in}}^{-4/3} (1-\Omega_{\sigma,{\rm in}})
\left( {\Gamma/H_{\rm dec}\over \Gamma/H_{\rm in}} \right)^{1/2}
& \quad {\rm for} \ y_{\rm dec}\ll
\Omega_{\sigma,{\rm in}}/(1-\Omega_{\sigma,{\rm in}})
\end{array}
\right.\,.
\end{equation}

\begin{figure}
\begin{center}
\includegraphics[width=80mm]{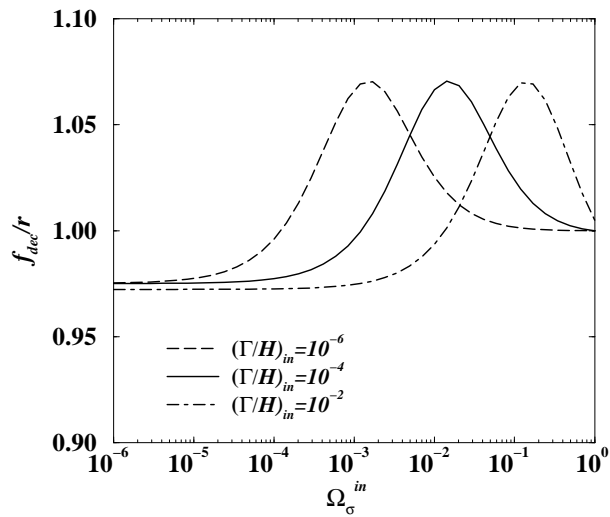}
\caption[sd]{\label{sd}
Comparison of full numerical solution for $r$ in Eq.~(\ref{transfer}) with
sudden-decay approximation, $f_{\rm dec}$ given in Eq.~(\ref{fdec}).}
\end{center}
\end{figure}

The sudden decay approximation is compared with numerical results
for the full equations~(\ref{zetaprime}) and
(\ref{zetasigmaprime}) in Figure~\ref{sd}.
The one free parameter in the sudden-decay approximation is the
particular value of $\Gamma/H_{\rm dec}$ chosen to characterise the
epoch of decay. Optimising the fit to the full numerical solutions
fixes $\Gamma/H_{\rm dec}\simeq1.4$, in line with our expectation that
$\Gamma/H_{\rm dec}$ should be of order unity. With this choice the
sudden-decay approximation is seen (Figure~\ref{sd}) to give a good
estimate for $r\simeq f_{\rm dec}$ (good to within 10\%).

\section{Conclusions}
\label{conc}

We have studied the evolution of large-scale curvature
perturbations for multiple interacting fluids in a linearly
perturbed FRW cosmology.
The curvature perturbation, $\zeta_\alpha$, on hypersurfaces of
uniform density for each fluid, Eq.~(\ref{zetaalpha}), provides a
gauge-invariant variable by which to study the large-scale
evolution. The total curvature perturbation, 
$\zeta$, is then a weighted sum, Eq.~(\ref{zetatot}), of the individual
$\zeta_\alpha$'s. For a non-interacting perfect fluid,
$\zeta_\alpha$ remains constant on large scales, independently of
perturbations in other fluids~\cite{WMLL}. More generally we have shown how
$\zeta_\alpha$ can change on large scales due to either an
intrinsic non-adiabatic pressure perturbation or non-adiabatic
energy transfer.

We can decompose an arbitrary energy transfer perturbation into two
parts:
\be
 \delta Q_\alpha = {\dot Q_\alpha \over \dot\rho_\alpha}
 \delta\rho_\alpha + \delta Q_{\alpha,{\rm intr}} \,,
\ee
where $\delta Q_{\alpha,{\rm intr}}$ is the (gauge-invariant)
intrinsic non-adiabatic energy transfer. The large-scale curvature
perturbation $\zeta_\alpha$ can change due to this intrinsic
non-adiabatic energy transfer, or due to a relative entropy
perturbation between fluids, $\delta Q_{\alpha,{\rm rel}}$ defined in
Eq.~(\ref{deltaQrelalpha}), which is proportional to $\zeta_\alpha-\zeta$.
For perturbations that obey the generalised adiabatic condition:
\be
\delta P_\alpha = c_\alpha^2 \delta\rho_\alpha \,,\ \
\delta Q_\alpha = {\dot Q_\alpha\over\dot\rho_\alpha} \delta\rho_\alpha \,\
\ 
{\rm and}\ \ 
\zeta_\alpha = \zeta \,,
\ee
the curvature perturbation, $\zeta_\alpha$, remains constant on large
scales.
If all the individual fluids obey this generalised adiabatic
condition, then the total curvature perturbation, $\zeta$, is
necessarily constant too.

Most previous analyses have adopted the variables of Kodama and
Sasaki~\cite{KS} who defined the relative entropy perturbation
between two fluids as 
\be 
\label{SKS}
{S}_{\alpha\beta} \equiv {\delta \rho_\alpha \over
\rho_\alpha+P_\alpha} - {\delta \rho_\beta \over
\rho_\beta+P_\beta} \,.
\ee
However this definition is only gauge-invariant in the absence of
energy transfer. As a result the evolution equations including energy
transfer are particularly unpleasant~\cite{KS,hama}. In particular it
is diffcult to show that entropy perturbations obey a homogeneous
evolution equation on large scales (i.e., adiabatic perturbations stay
adiabatic on large scales) in the way that was recently shown for
multiple interacting scalar fields~\cite{chris}. Although it has been
be argued on very general grounds that this must be the
case~\cite{WMLL}, this fundamental result has not previously been
explicitly demonstrated in treatments of interacting fluids.

We have used the correct gauge-invariant generalisation of
(\ref{SKS}), allowing for energy transfer,
\be
 \S_{\alpha\beta} \equiv 3(\zeta_\alpha-\zeta_\beta)
 \,,
\ee
which describes the relative displacement between the two
hypersurfaces of uniform density defined with respect to the two
fluids. This reduces to (\ref{SKS}) in the case of no energy
transfer. It allows us to demonstrate that the evolution of the
large-scale entropy perturbation, Eq.~(\ref{S_evol}) is sourced
only by entropy perturbations and not sourced by the total
curvature perturbation, $\zeta$. Thus the integrated evolution on
large scales, even when we include energy transfer, can be
schematically represented by the linear transfer matrix:
\be
 \label{trans2}
 \left(
\begin{array}{c}
 \zeta \\ \S
\end{array}
\right)_{\rm out}
 =
\left(
 \begin{array}{cc}
1 & T_{\zeta\S} \\
0 & T_{\S\S}
 \end{array}
\right) \left(
\begin{array}{c}
\zeta \\ \S
\end{array}
 \right)_{\rm in} \,.
\ee

We have applied our formalism to study the evolution of curvature
perturbations in the curvaton scenario where an initially
isocurvature (non-adiabatic) perturbation in the curvaton field is
transferred to the radiation fluid when the curvaton eventually
decays. The decay of the curvaton represents a non-adiabatic
energy source for the radiation fluid. We have numerically solved
the coupled evolution equations to determine the resulting
curvature perturbation, $\zeta$, for an initial entropy
perturbation. Thus we have calculated the transfer coefficient
$T_{\zeta\S}$ in Eq.~(\ref{trans2}) for different parameter values
of the background models. We compared our results with
semi-analytic estimates based on the ``sudden-decay''
approximation \cite{curvaton,LUW} where the fluids are assumed to
be non-interacting up until a fixed decay time. The sudden-decay
approximation is shown to give a good fit to the full result
(within 10\%) for a suitable choice of fitting parameter.

In this two-fluid realisation of the curvaton scenario the
interaction between the fluids leads to the relative entropy
decaying to zero at late times, $T_{\S\S}=0$ in
Eq.~(\ref{trans2}), leaving a purely adiabatic curvature
perturbation. Our formalism can also be applied to cosmological
models including other cosmological fluids such as baryons, CDM or
neutrinos, in which case it should be possible to calculate the
amplitude of any residual isocurvature perturbations that may
survive after curvaton decay in different variations of the
curvaton scenario.

\acknowledgments

The authors are grateful to David Lyth, Misao Sasaki and Filippo
Vernizzi for useful comments. 
This work was supported by PPARC grant \emph{PPA/G/S/2000/00115}.
KM is supported by a Marie Curie Fellowship
under the contract number \emph{HPMF-CT-2000-00981}.
DW is supported by the Royal Society.

{}

\end{document}